\documentclass[aps,prl,superscriptaddress,reprint,showpacs,longbibliography,amsmath,amssymb,nofootinbib]{revtex4-2}

\usepackage{amsmath}
\usepackage{amssymb}
\usepackage{amsfonts}
\usepackage{latexsym}
\usepackage{graphicx}
\usepackage{color}
\usepackage[colorlinks,linkcolor=blue,citecolor=blue,urlcolor=blue]{hyperref}
\usepackage{cleveref}
\usepackage{microtype}
\usepackage{lipsum}
\usepackage{tikz}
\usepackage{lmodern}
\newcommand\encircle[1]{%
  \protect\tikz[baseline=(X.base)] 
    \protect\node (X) [draw, shape=circle, inner sep=-1.2pt, scale=0.9] {\strut #1};}

\usepackage{xcolor}

\usepackage{glossaries}
\setacronymstyle{long-short}
\DeclareUnicodeCharacter{2212}{-}
\glsresetall 

\begin{document}

\title{Emergent cavity junction around metal-on-graphene contacts}
\author{Yuhao Zhao}
\thanks{Y.Z. and M.K. contributed equally to this work.}
\affiliation{Institute for Theoretical physics, ETH Zürich, 8093 Zürich, Switzerland}

\author{Maëlle Kapfer}
\thanks{Y.Z. and M.K. contributed equally to this work.}
\affiliation{Department of Physics, Columbia University, 10027 New York, NY, USA}

\author{Kenji~Watanabe}
\affiliation{National Institute for Materials Science, 1-1 Namiki, Tsukuba, 305-0044, Japan}

\author{Takashi~Taniguchi}
\affiliation{National Institute for Materials Science, 1-1 Namiki, Tsukuba, 305-0044, Japan}

\author{Oded~Zilberberg}
\affiliation{Department of Physics, University of Konstanz, 78464 Konstanz, Germany}

\author{Bjarke~S.~Jessen}
\affiliation{Department of Physics, Technical University of Denmark, 2800 Copenhagen, Denmark}









\maketitle

\textbf{Harnessing graphene devices for applications relies on a comprehensive understanding of how to interact with them. Specifically, scattering processes at the interface with metallic contacts can induce reproducible abnormalities in measurements. Here, we report on emergent transport signatures appearing when contacting sub-micrometer high-quality metallic top contacts to graphene. Using electrostatic simulations and first-principle calculations, we reveal their origin: the contact induces an n-doped radial cavity around it, which is cooperatively defined by the metal-induced electrostatic potential and Klein tunneling. This intricate mechanism leads to secondary resistance peaks as a function of graphene doping that decreases with increasing contact size. Interestingly, in the presence of a perpendicular magnetic field, the cavity spawns a distinct set of Landau levels that interferes with the Landau fan emanating from the graphene bulk. Essentially, an emergent 'second bulk' forms around the contact, as a result of the interplay between the magnetic field and the contact-induced electrostatic potential. The interplay between the intrinsic and emergent bulks leads to direct observation of bulk-boundary correspondence in our experiments. Our work unveils the microscopic mechanisms manifesting at metal-graphene interfaces, opening new avenues for understanding and devising graphene-based electronic devices.}

\glsresetall 


Since the isolation of monolayer graphene, interfacing two-dimensional (2D) materials with the three-dimensional world has been an ongoing research field of intense interest~\cite{lee2008contact, PhysRevB.79.245430, PhysRevLett.101.026803, PhysRevLett.104.076807, matsuda2010contact, wang2013one, leong2014low, 10.1063/1.3077021}.
The ubiquity of this interface, typically deposited metal on the graphene, along with the unique properties of graphene, poses an intriguing challenge and a departure from conventional semiconductor physics.
Ongoing efforts mainly concern 2D semiconductors, such as MoS$_2$ and WSe$_2$~\cite{allain2015electrical, liu2016van, kim2017fermi}, as the metal-on-graphene (MG) interface is largely considered to be a solved problem~\cite{xia2011origins, 6109352, wu2012quantum, cusati2017electrical}.
In particular, the MG interface is usually understood using a relatively simple modification of graphene's work function alongside electrostatic considerations~\cite{huard_evidence_2008,allain_electrical_2015,matsumoto_frontiers_2015}: the graphene under the metal has its potential fixed to the work function of the metal and the potential gradually returns to the intrinsic graphene level away from the MG interface. 
The resulting doping gradient and the carrier tunability of graphene away from the metal leads to the formation of pn and pp’ junctions, which in turn explains a characteristic asymmetry in two-terminal resistance measurements~\cite{xia2011origins,bartolomeo_graphene_2015}.	However, other features often show up, such as extra resistance peaks, which are considered as unavoidable imperfections and are largely ignored as they are absent in four-terminal measurements, where effects at and around the metal contacts are excluded~\cite{Datta_1995}. Furthermore, such extra resistance features are often, and incorrectly, attributed to the graphene directly under the metal contacts~\cite{song_determination_2012}.

  \begin{figure*}[th!]
		\centering
		\includegraphics{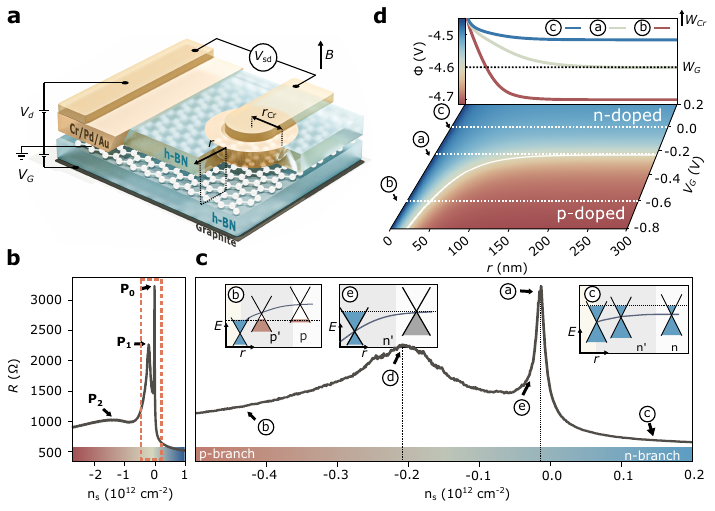}
		\caption{\textbf{Emergent $pn$-junction at a metal-graphene interface.} \textbf{a.} Graphene is encapsulated in \textit{h}BN and placed on a graphite back gate. Top contacts (Cr/Pd/Au) are gently deposited directly on the graphene after selectively etching through the top \textit{h}BN layer. \textbf{b.}  Two-terminal resistance between the large contact (left) and a small dot contact (right) as a function of the carrier density $n_s$. We observe three resistance peaks $\mathbf{P}_0$-$\mathbf{P}_2$. \textbf{c}. Zoom in on the dashed box in \textbf{b}: \encircle{a}-\encircle{e} mark significant features, discussed in the text. Insets: corresponding radial interpolation of the work function of graphene from the contact to the bulk. \textbf{d.}  The electrostatic potential around the contact is simulated using a finite-element method, see Methods. The obtained potential is plotted as a function of the carrier density, $n_s\propto V_G$, and the radial distance, $r$, starting from the contact edge. Presented linecuts at \encircle{a}-\encircle{c} mark scenarios where the pristine graphene is at the neutral point (white), $p$-doped (red), and $n$-doped (blue), respectively. }
        \label{fig: setup and zero-resistance}
\end{figure*}
\begin{figure*}[th]
		\centering
		\includegraphics{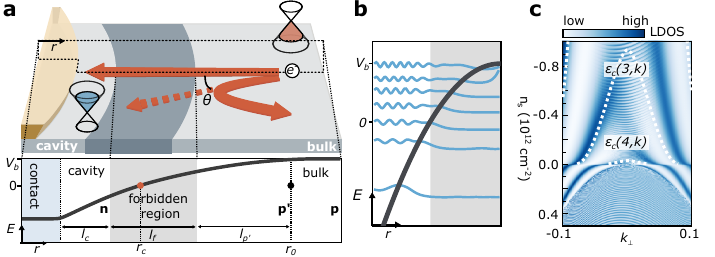}
		\caption{\textbf{Doping-dependent cavity formation and energy quantization.} \textbf{a.} Upper panel: Real space illustration of electron transport into the contact when the graphene bulk carrier density $n_s$ is in the p-branch. Dopings are marked by the filling of the Dirac cone. Red arrows depict that Klein tunneling is relevant only at $k_\perp\approx 0$. Lower panel: corresponding approximate parabolic function potential profile around the contact, see Fig.~\ref{fig: setup and zero-resistance}\textbf{d}. The thick black line marks the position of the Dirac point along the radial direction. The interpolating on-site potential leads to the formation of three regions in the graphene (i) $n$-doped cavity (blue-filled cone) of width $l_\textrm{c}$, (ii) energetically-forbidden region (dark gray) defining the $pn$-junction of width $l_\textrm{f}$, and (iii) the $p$-doped graphene bulk. The latter is split into two parts $p'$ and $p$, until the bulk doping is stabilized. \textbf{b.} Sketch of bound states with discretized energy levels in the emergent cavity as a function of $r$. \textbf{c.}  Local density of states (LDOS) of the bound states (up to $r\approx50\,\textrm{nm}$) as a function of the transverse momentum $k_\perp$ and doping level $n_s$, see Methods. Fine oscillations are due to finite-size effects in the numerics. Modes around $k_\perp\approx 0$ penetrate the cavity due to Klein tunneling. The white dashed line marks the analytically obtained spectrum of the cavity mode $\epsilon_c(m,k_\perp)$, see Methods.}
    \label{fig: 1D system, resistance with varying sites}
\end{figure*}

Here, we present findings challenging this long-held paradigm, revealing an unexpected and extremely rich structure within the MG interface, which becomes particularly visible when examining contacts with sub-micrometer dimensions. To demonstrate this, we perform our experiments in the simplest possible setup: a single high-quality contact on graphene connected to a larger reservoir contact. Using a combination of magnetotransport, electrostatic simulations, and \textit{ab initio} calculations, we explain the origin of the emergent structure in this simple disc-shaped contact geometry. Our key discoveries include an intricate potential landscape near the contact that gives rise to quantized cavity states and, under a magnetic field, the formation of a distinct emergent ‘second bulk’ through the formation of an emergent “wedding-cake”-like renormalization of the potential landscape~\cite{gutierrez_interaction-driven_2018}. This emergent bulk allows us to directly observe topological edge states that form at the emerging interface between the two bulks. Our results underscore the critical need for precise knowledge of sub-micrometer graphene interfaces, a cornerstone for scaling down electronic components towards spintronics~\cite{avsar2020colloquium} and electron optics applications~\cite{hawkes2022principles,zhao_electron_2023}. This understanding is particularly important as new generations of ever-improving two-dimensional material platforms unravel richer physics~\cite{bolotin2009observation, wang2013one, zibrov2017tunable, rhodes2019disorder}.

\section*{Setup and transport measurement}
Our samples consist of monolayer graphene encapsulated within \textit{h}BN supported by a graphite back gate, allowing modulation of the carrier density in the graphene layer, see Fig.~\ref{fig: setup and zero-resistance}\textbf{a}. A fluorine-based dry etch (see Methods) is used to selectively etch the \textit{h}BN to define top contacts. A metal stack of Cr/Pd/Au is then deposited on the exposed graphene resulting in a gentle defluorination process, such that a high-quality physisorbed metallic layer is formed directly on the graphene~\cite{son2018atomically, kretz2018atomistic, jessen2019lithographic}. Source contacts are circular dots with a radius, $r_{\text{Cr}}$, down to 100\,nm, while drain contacts are relatively large rectangles with areas exceeding 10$\,\mu\text{m}^2$, allowing us to focus on the behavior of individual contacts. Due to the etch-angle of \textit{h}BN, a frustum-shaped top contact is realized, where the radius of the top surface is slightly larger than that of the bottom surface (see Methods). We note that the distance between contacts eliminates unwanted coherence between source and drain contacts. 

We first measure the two-terminal resistance, $R$, as a function of the carrier density, $n_s$, by varying the back-gate voltage bias $V_G$, see Fig.~\ref{fig: setup and zero-resistance}\textbf{b}. 
We use bias voltages around 100\,$\mu$V to avoid any unwanted charge-accumulation on the substrate~\cite{chiu_controllable_2010}.
Scanning over $n_s$ from $-3\times 10\,\mathrm{cm}^{-2}$ to $1\times10\,\mathrm{cm}^{-2}$, we observe a sharp resistance peak $\mathbf{P}_0$ at $n_s = 0\,\text{cm}^{-2}$, and two smooth resistance peaks $\mathbf{P}_1$ and $\mathbf{P}_2$ at $n_s = -0.2\,\text{cm}^{-2}$ and $n_s = -1.2\,\text{cm}^{-2}$, respectively. Narrowing our focus to the $n_s$ regime around $\mathbf{P}_0$, we attribute the sharp peak $\mathbf{P}_0$ to the Dirac point of the bulk graphene, marked by \encircle{a} in Fig.~\ref{fig: setup and zero-resistance}\textbf{c}. As $|n_s|$ increases (marked by \encircle{e}), $R$ rapidly drops due to the increasing density of states of the bulk graphene, $\mathrm{DOS}\propto\sqrt{n_s}$.  Asymptotically away from the first peak, $R$ exhibits an asymmetric imbalance in the $p$- and $n$-doped bulk regions, dubbed  $p$- and $n$-branch and marked by \encircle{b} and \encircle{c}, respectively. 
This is caused by the pinned potential under the metal contact, which effectively leads to the formation of a $pn$ junction between the $p$-doped bulk graphene and an $n$-doped region around the contact; the asymmetry in $R$ is then due to Klein tunneling filtering of conduction channels~\cite{cheianov_selective_2006,huard_evidence_2008,xia2011origins}. The appearance of a single secondary peak is conventionally attributed to the gap opening due to the g/\textit{h}BN misalignment~\cite{hunt_massive_2013,wolf_substrate-induced_2018}, or to the pinned Dirac point of the graphene under the contact~\cite{song_determination_2012}. However, we observe that the soft peak $\mathbf{P}_1$ (marked by \encircle{d}) appears too close to the main peak $\mathbf{P}_0$, rendering the g/\textit{h}BN misalignment explanation inapplicable. This, and the appearance of the additional smooth peak $\mathbf{P}_2$, points towards previously overlooked mechanisms, which go beyond standard graphene $pn$-junction physics.

\section*{Finite element simulation}
To gain insight into the microscopic origin of these resistance features, we first calculate, using a finite-element simulation, the electrostatic potential profile of the graphene, $\Phi$, around the metal contact, see Fig.~\ref{fig: setup and zero-resistance}\textbf{d} and Methods. The convex edge of the contact suppresses interference effects and allows us to isolate the subtle physics within the contact/bulk region~\cite{shytov_klein_2008}. We employ boundary conditions fixed by the work function of chromium $W_\mathrm{Cr}=4.3\,\text{V}$~\cite{song_determination_2012} and graphene $W_\mathrm{G}=4.6\,\text{V} + \sqrt{\pi v_F^2\hbar^2 C}\sqrt{V_G}$~\cite{yu_tuning_2009,shi_work_2010,ziegler_variation_2011}, where $v_F$ is the Fermi velocity; $\hbar$ is the reduced Planck constant and the geometric capacitance $C=2.72\times 10^{15}\,\mathrm{V}^{-1}\mathrm{cm}^{-2}$. Under the contact we assume that the potential of the graphene is pinned to $W_\mathrm{Cr}$ regardless of $V_G$~\cite{song_determination_2012,matsumoto_frontiers_2015}, i.e., the graphene is $n$-doped there. Radially away from the contact edge, the electrostatic potential continuously transitions from the pinned value to the work function of the bulk graphene, set by $V_G$. In the $p$-branch, a $pn$-junction emerges $\sim30$--$150$\,nm away from the contact's edge, marked by a solid white line in Fig.~\ref{fig: setup and zero-resistance}\textbf{d}.
The emergence of a $pn$-junction as a function of $V_G$ is in agreement with the asymmetry in resistance observed in Fig.~\ref{fig: setup and zero-resistance}\textbf{c}. Moreover, to verify the obtained distance between the $pn$-junction and the contact's edge, which is seemingly shorter compared with previous research~\cite{mueller_role_2009,cusati2017electrical}, we perform the same measurement (see Extended data Fig.~6) with two contacts that are distanced $350\,\mathrm{nm}$ from each other. We observe the asymmetry in resistance, suggesting the existence of a very narrow $pn$-junction close to the contact's edge.

\begin{figure*}[th]
		\centering
		\includegraphics{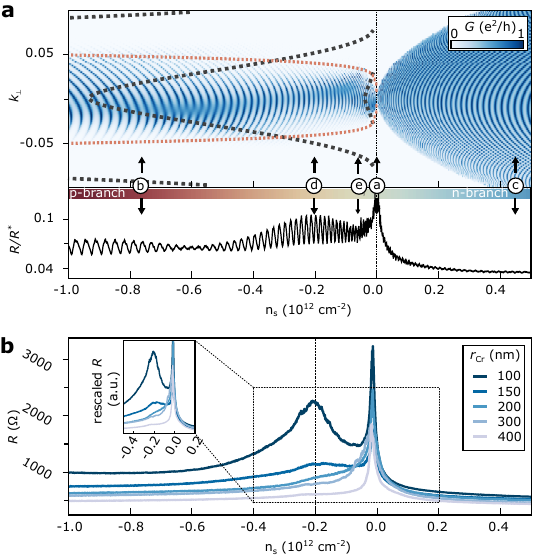}
		\caption{\textbf{Cavity states modulated resistance and the impact of contact size.} \textbf{a.} Upper panel: Conductance [cf.~Eq.~\eqref{eq: conductance}] as a function of momentum $k_\perp$ and carrier density $n_s$. Fine oscillations are due to the finite size of the numerical solution. Red dashed lines encircle the regime of high-tunneling over the forbidden region where $t_{pn}>0.5$, i.e., where Klein-tunneling is dominant (see Methods). The dispersion of the cavity modes $\epsilon_c(m,k_\perp)$ is marked by gray dashed lines, cf.~Fig.~\ref{fig: 1D system, resistance with varying sites}\textbf{c}. Lower panel: The corresponding resistance as a function of $n_s$ in units of $R^*=h/(4e^2)$. A soft resistance peak around $n_s=-0.2 \times 10^{12} \text{cm}^{-2}$ appears with small oscillations due to finite-size numerical artifacts.
        \textbf{b.} Experimentally measured resistance as a function of the carrier density $n_s$ for different dot contacts with radii ranging from $100\ \text{nm}$ to $400\ \text{nm}$. Inset: Zoom-in plot of $R$ in the marked area in \textbf{a}. The resistance is re-scaled relative to the $R$ measured with $r_\mathrm{Cr}=100\,\mathrm{nm}$ at $n_s=1.0 \times 10^{12} \text{cm}^{-2}$.}
    \label{fig: 1D system, resistance with varying sites 2}
\end{figure*}

\section*{Cavity forming around the contact}
For $n_s<-0.1\,\text{cm}^{-2}$, the $n$-doped region between the contact's edge and the charge neutrality point (CNP) can become very narrow with a radial width down to $\sim30\,\text{nm}$, see Fig.~\ref{fig: setup and zero-resistance}\textbf{d}. To explore the prospective quantization of modes in this narrow region, we write the graphene Hamiltonian around the contact using a mixed position and momentum representation, i.e., using a finite radial extent $\hat{r}$ and transverse momentum $k_\perp\equiv|\mathbf{k}|\sin(\theta)$, where $\theta$ is the incident angle at the interface, and $\mathbf{k}$ is the 2D momentum, see Fig.~\ref{fig: 1D system, resistance with varying sites}\textbf{a}. We introduce the interpolating electrostatic potential as an inverse-parabola on-site term $V(r,V_b)$, characterized by two parameters $V_0=V(0,V_b)$ and $\alpha=\partial^{2}_{r}V$ (see Methods). This on-site term extends between the contact and the graphene bulk dopings, while the latter is denoted by $V_b$, and experimentally tuned by $V_G$. It engenders a quasi-triangular radial cavity around the contact with bound modes, see Fig.~\ref{fig: 1D system, resistance with varying sites}\textbf{b}. The cavity width is $k_\perp$-dependent: consider a state propagating outwards from the contact along $\hat{r}$; it backscatters when the on-site energy is equal to the kinetic energy in the transverse direction, i.e., we have a classical turning point when $|V_b|=v_F|k_\perp|$. Similarly, another classical turning point manifests for an inbound electron propagating towards the contact. The resulting two classical turning points thus determine an effective $k_\perp$-dependent (i) width for the cavity $l_c= \sqrt{|V_b-V_0|/\alpha}-\sqrt{(V_b+|k_\perp|v_F)/\alpha}$, as well as, (ii) a classically-forbidden region of width $l_f$ between the two turning points, see lower panel of Fig.~\ref{fig: 1D system, resistance with varying sites}\textbf{a}. We observe that both $l_c$ and $l_f$ decrease with increasing $n_s$. The former entails that the energy spacing between the cavity levels increases with $n_s$. The latter implies that the tunneling barrier into the bulk becomes narrower with increasing $n_s$. We verify this prediction using a Green's function method, by which we numerically obtain the local density of states (LDOS) spectrum in the cavity, see Figs.~\ref{fig: 1D system, resistance with varying sites}\textbf{b} and \textbf{c}. Crucially, we observe a discrete set of energy levels that depend on $k_\perp$ alongside contributions from Klein tunneling at $k_\perp\approx 0$. An analytically obtained spectrum $\varepsilon_c(m,k_\perp)$ (see Methods) fits well with the result. In the $n$-branch, the absence of a forbidden region allows states near the contact to strongly hybridize with the $n$-doped graphene bulk, leading to a continuous dispersion. 

We move to calculate the conductance $G_{k_\perp}$ between the contact and the bulk (see 
Figure~\ref{fig: 1D system, resistance with varying sites 2}\textbf{a})
\begin{equation}\label{eq: conductance}
    G_{k_\perp}(V_b)\propto|t_{pn}(V_b,k_\perp)|^2\rho_{k_\perp}(V_b)\,,
\end{equation}
 where $t_{pn}$ is the tunneling amplitude through the forbidden region and $\rho_{k_\perp}(V_b)$ is a $k_\perp$-dependent tunneling density of states; the latter depends on both the density of states in the bulk and in the cavity. We observe strikingly different behavior in the $p$- and $n$-branches. In the $n$-branch, where no $pn$-junction nor cavity form, $G_{k_\perp}$ exhibits a parabolic shape as a function of $k_\perp$ and $n_s$, indicating that transport is dominated by the density of states of the graphene bulk. In the $p$-branch, instead, $G_{k_\perp}$ is restricted to small $k_\perp$, due to filtering of $t_{pn}$ via Klein tunneling through the forbidden region. Crucially, whenever the cavity modes coincide with the allowed transport window, the $G_{k_\perp}$ intensity increases, indicating a modification of the density of scattering states $\rho_{k_\perp}(V_b)$.
 
The total resistance is obtained as $R=[\sum_{k_\perp}G(k_\perp)]^{-1}$, see bottom panel of Fig.~\ref{fig: 1D system, resistance with varying sites 2}\textbf{a}, which we compare with Fig.~\ref{fig: setup and zero-resistance}\textbf{c}. Starting from the neutral point \encircle{a}, the resistance is high due to the absence of states in the bulk. As $n_s$ increases towards \encircle{c} in the $n$-branch, the resistance drops due to the increasing DOS in the bulk. Conversely, as $n_s$ decreases towards \encircle{e}, the resistance is initially reduced as a dense spectrum of cavity states is available at small $k_\perp$. The $\mathbf{P}_1$ peak at $n_s\approx -0.2\times 10^{12}\,\text{cm}^{-2}$, marked as \encircle{d}, appears when the cavity modes are off-resonant with the Klein-tunneling allowed transport channels. Finally, at \encircle{b}, although the cavity states found at small $k_\perp$ can assist the tunneling to the contact, their impact is less pronounced as the narrow cavity and forbidden region allow for direct co-tunneling between the bulk and the contact. Nevertheless, when $n_s$ continues to decrease towards the deeply $p$-doped regime (see Extended data Fig.~4), a second smooth peak is observed where the cavity modes are once more off-resonant with the allowed transport channels, thus, yielding a similar feature as  $\mathbf{P}_2$.

\subsection*{Scaling with contact size}

Our effective cavity model suggests that the secondary peak originates from the reduced DOS in the emergent confined cavity around the contact, and not from the graphene directly under the metal. As such, by increasing the contact size, we expect an increase in the DOS due to added transverse momentum modes, and hence a reduction in the peak's height. To probe our prediction, we repeat the two-terminal resistance measurement using dot contacts with radii $r_\mathrm{Cr}$ ranging from $100\ \text{nm}$ to $400\ \text{nm}$, see Fig.~\ref{fig: 1D system, resistance with varying sites 2}\textbf{b}. As the size of the metal contact increases, more transport channels participate, and the overall $R$ is reduced. Nonetheless, the peak is observed for all contact sizes at the same carrier density $n_s\approx-0.2\ \times 10^{12}\ \text{cm}^{-2}$, showing that the secondary resistance peak indeed depends only on the potential environment created by the metal contact. To better isolate the peaks's qualitative $R$ reduction from the increased cavity DOS on top of the $R$ reduction due to the contact size, we rescale all measured plots relative to the $R$ in the $n$-region, where no cavity forms, see inset of Fig.~\ref{fig: 1D system, resistance with varying sites 2}\textbf{b}. The cavity remains visible while showing a clear reduction with the cavity diameter.

\begin{figure*}[!t]
		\centering
		\includegraphics[width=\textwidth]{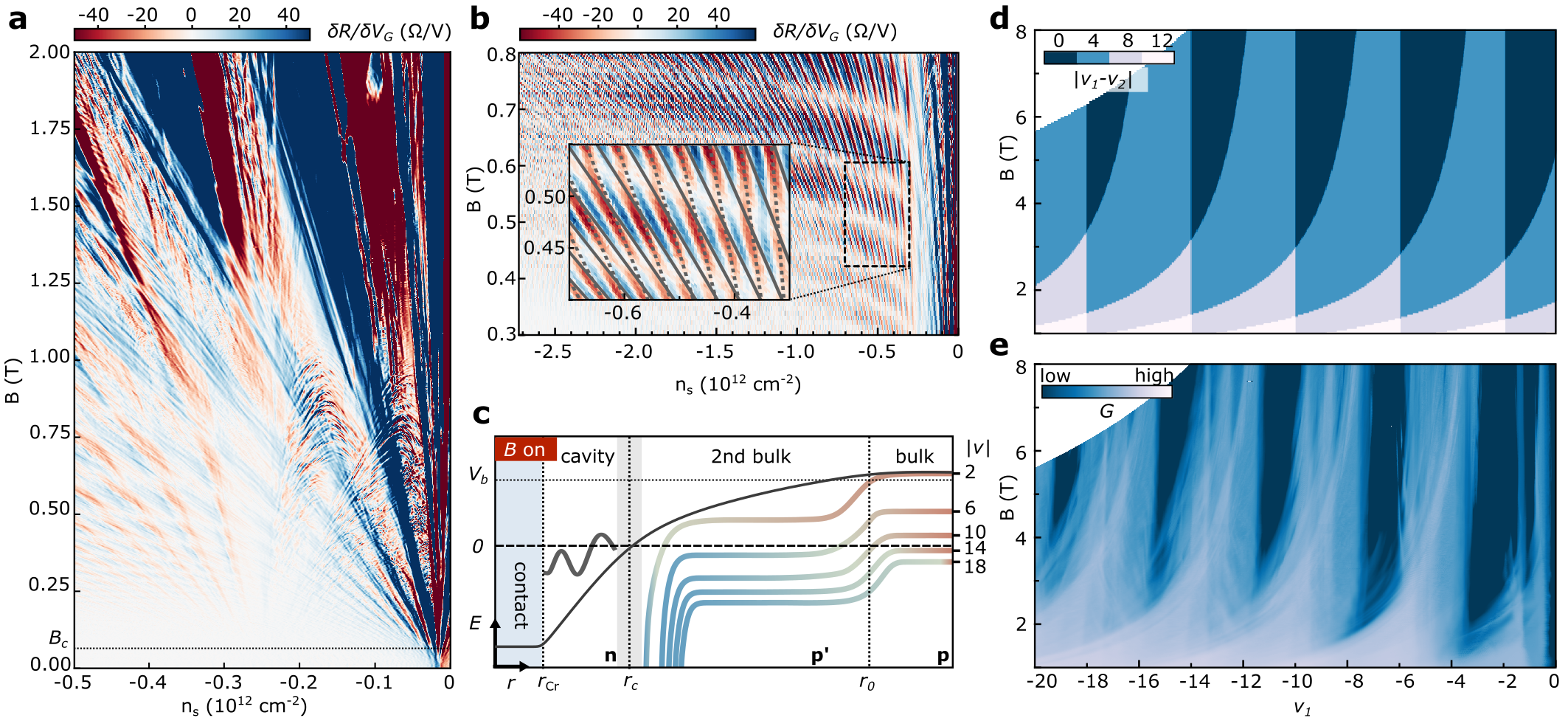}
    ~		\caption{\textbf{Formation of a second bulk with a perpendicular magnetic field.} \textbf{a.} Differential resistance at low-field and small $n_s$. Two distinct sets of Landau Fans are observed. \textbf{b.}  Differential resistance for higher density values with inset highlighting phase slips at the crossing of the Landau levels. \textbf{c.} Sketch of ``wedding cake'' potential profile around the contact forming under a perpendicular magnetic field. The smooth interpolating potential (black solid line) is re-normalized and Landau levels appear along a step-like energy profile (colored lines), cf.~Ref.~\citep{gutierrez_interaction-driven_2018}.  \textbf{d.}  The difference between the fillings of the two bulks $|\nu_1-\nu_2|$,  as a function of the filling factor of the graphene bulk $\nu_1$ and the magnetic field. \textbf{e.}  The longitudinal conductance $G$ measured as a function of $\nu_1$ and $B$. The conductance exhibits a triangular-shaped pattern resembling the $|\nu_1-\nu_2|$ from \textbf{d}. We observe further splitting between the Landau fans of the secondary bulk.}
    \label{fig: dual bulk formation}
\end{figure*}
\section*{Dual Landau Fans}

Subjecting the MG system to a perpendicular magnetic field reveals further intricate details about its electronic properties. In Fig.~\ref{fig: dual bulk formation}\textbf{a}, we present the measured differential resistance ($\delta R/\delta V_G$) as a function of the magnetic field, $B$, and carrier density, $n_s$. While an expected Landau fan (LF) pattern emerges out of graphene's CNP, interestingly, a second LF appears near $n_s \approx -0.24 \times 10^{12}\,\text{cm}^{-2}$, i.e., coinciding closely with the first cavity-induced resistance peak, $\mathbf{P}_1$. Probing a larger density range (Fig.~\ref{fig: dual bulk formation}\textbf{b}), apparent phase slips at the intersections of the two LFs are evident. While the primary LF is associated with the bulk graphene, the origin of the second LF is unconventional: (i) we cannot attribute it to the recoil band forming from the misalignment between graphene and \textit{h}BN ~\cite{hunt_massive_2013,wolf_substrate-induced_2018}, as the observed $n_s \approx -0.24 \times 10^{12}\,\text{cm}^{-2}$ is too small compared with the typical case related to such recoil; additionally, $\mathbf{P}_1$ and the source of the emanating secondary LF occurs at a similar value of $n_s$ across all our MG samples, regardless of g/\textit{h}BN alignment (not shown); (ii) the second LF manifests only beyond a finite field strength ($B_c \approx 0.1\,\mathrm{T}$), which may hint towards a Fock-Darwin cascade of states within the emergent cavity~\cite{chen_fock-darwin_2007}; we rule out this explanation as the cavity length ($l_c \approx 20\,\text{nm}$) is smaller than the magnetic length ($l_{B_c} \approx 80\,\text{nm}$).
\section*{Direct view of the bulk-boundary correspondence}
We attribute the second LF to a secondary bulk forming between the cavity and the bulk graphene. Without a magnetic field, a smoothly-interpolating $p'$-doped region exists between the bulk and the cavity ($p'\neq p$), cf. Fig~\ref{fig: 1D system, resistance with varying sites 2}\textbf{a}. When the magnetic field is applied, electrons can be trapped in the $p'$-doped region if their cyclotron movement is confined within the region’s scale, i.e., when  $l_{B_c}\approx l_{p'}$~\cite{gu_collapse_2011}. Similar to LLs, these electrons fall into highly degenerate energy levels, where Coulomb interaction can renormalize the potential profile~\cite{gutierrez_interaction-driven_2018}. Consequently, the potential in the $p'$-doped region flattens out such that a second bulk effectively forms and the overall potential exhibits steps akin to a "wedding cake", see Fig.~\ref{fig: dual bulk formation}\textbf{c}. The energetic offset at the emergent bulk implies a secondary LF emanating from a carrier density $n_s$, where the graphene bulk is $p$-doped, i.e., at $n_s<0\,\mathrm{cm}^{-2}$. Furthermore, this offset implies that the Landau level filling factors, $\nu=4(n+1/2)$, differ between two bulks. We are hence in a scenario where the quantum Hall effect's Chern numbers $n\in\mathbb{N}$ vary radially away from the MG contact between the two differently-doped bulks. Due to the different topology between the two bulks, at the interface between them ($r\sim r_0$), $|\nu_1-\nu_2|$ edge modes appear~\cite{hasan2010colloquium}, where $\nu_{1,2}$ are the filling factors in the main and the secondary bulks, respectively, see Fig.~\ref{fig: dual bulk formation}\textbf{d}. In Fig.~\ref{fig: dual bulk formation}\textbf{c}, we illustrate this bulk-boundary correspondence~\cite{hatsugai_chern_1993}, i.e., the continuous change in bulk filling leads to $|\nu_1-\nu_2|$- intersections of LLs with the Fermi level ($E=0$) at the interface between the two bulks ($r\sim r_0$), thus forming edge states. As the emergent edge states appear near the MG contact, we expect the conductance to be dominated by the presence of these transport channels. We confirm this model prediction by measuring within the same parameter range of Fig.~\ref{fig: dual bulk formation}\textbf{d}, see  Fig.~\ref{fig: dual bulk formation}\textbf{e}. We observe a strong agreement between the measured conductance $G$ and the residual filling $|\nu_1-\nu_2|$. Furthermore, we observe splitting of the LLs that appear as a nested set of triangular shapes; we attribute these splittings to the lifting of the spin-valley degeneracy in the secondary bulk.

\section*{Conclusion and outlook}

Our work reveals a surprising complexity in the seemingly simple system of a metal contact on graphene. Contrary to the conventional picture of mere doping, the contact creates an intricate potential landscape that gives rise to quantized cavity states. Under a magnetic field, this landscape further evolves, spawning a distinct emergent bulk with its own set of Landau levels. This second bulk, arising from the interplay between the metal-induced potential and electron-electron interactions, leads to the direct observation of the bulk-boundary correspondence through the formation of topological edge states at the interface between the two bulks.
The emergence of these unexpected features – quantized cavity states, a secondary bulk, and emergent topological edge states – within a single graphene sheet challenges our understanding of metal-graphene interfaces. It suggests that even seemingly well-understood contact phenomena can harbor hidden complexities with significant implications for device physics. Our findings open new avenues for exploring and controlling these features, potentially leading to novel functionalities in graphene-based electronic devices. For instance, the ability to generate confined regions with distinct topological properties could find uses in the emergent field of topological electronics. Furthermore, the direct observation of topological edge states at emergent bulk boundaries offers a new platform for investigating the fundamental properties of such topological transport in condensed matter systems.


\section*{Acknowledgments}

O.Z. and Y.Z. acknowledge support by the Swiss National Science Foundation, the Deutsche Forschungsgemeinschaft (DFG) through Project No. 449653034, and Eidgenössische Technische Hochschule Zürich research Grant No. ETH-28 23-1.
B.S.J. and M.K. acknowledge support from the Center for Programmable Quantum Materials (Pro-QM), an Energy Frontier Research Center funded by the US Department of Energy (DOE), Office of Science, Basic Energy Sciences (BES), under award DE-SC0019443. 
B.S.J. acknowledges support from the Novo Nordisk Foundation grant NNF23OC0084494, BioNwire.
We thank Peter Bøggild and Cory R. Dean for useful discussions and feedback on the manuscript.

\section*{Author Contributions}
B.S.J. and M.K. prepared the samples and performed the transport measurements.
Y.Z. performed finite element simulations, cavity simulations, and developed the analytical model. 
O.Z. developed the analytical model and provided technical support for the numeric simulations.
B.S.J., M.K., and Y.Z. analyzed the data.
K.W. and T.T. grew and provided the \textit{h}BN crystals.
Y.Z., B.S.J., M.K., and O.Z. wrote the paper with input from all authors. 
\section*{Competing Interests}
The authors declare no competing interests. 


\clearpage
\newpage
\section*{Methods}

\subsection*{Device fabrication}
Graphene and \textit{h}BN are exfoliated on a SiO$_{2}$/Si substrate, and single-layer graphene is automatically identified using its contrast relative to the substrate \cite{jessen2018quantitative}. The graphene heterostructures are fabricated with a standard dry-stacking method \cite{wang2013one, pizzocchero_hot_2016,purdie_cleaning_2018} using a polycarbonate-coated polydimethylsiloxane stamp. We sequentially pick up and intentionally misalign \textit{h}BN ($\sim30\,$nm), monolayer graphene, \textit{h}BN ($\sim30\,$nm), and deposit these on a $\sim5\,$nm thick graphite flake. Layers are picked up by raising our substrate temperature to about 110$^{\circ}$C during stamp contact, and dropped onto the graphite flake by melting the polycarbonate at around 190$^{\circ}$C. Subsequently, we use electron beam lithography to define contacts in a two-step process. First, the top \textit{h}BN is gently etched using a highly selective fluorine-based dry etch, based on either CF$_{4}$ or SF$_{6}$, to expose the graphene. Crucially, this process does not etch through the graphene (Extended Data Fig.~1). This is directly followed by a metal deposition of Cr/Pd/Au ($\sim3/15/60\,$nm) to contact the graphene. A second lithography step is done to draw electrical leads to the patterned nanostructures. All metals are deposited at pressures of $\sim10^{-8}\,$Torr and rates of 0.3\,Å/s, 1\,Å/s, and 1\,Å/s for Cr, Pd, and Au, respectively.

\subsection*{Measurements}
Transport measurements were performed in a Bluefors dry dilution cryostat achieving a temperature at the mixing chamber of 10\,mK. The electrical resistance was measured in a two-terminal configuration using Stanford Research SR860 lock-in amplifiers while sourcing 100$\,\mu$V at a reference frequency of 17.777\,Hz to efficiently reject noise. Commercial Q-Devil filters are placed on the mixing chamber of the cryostat adding a 5k\,$\Omega$ impedance in series with the device which we subtract in the presented data.

\subsection*{Finite element calculation}
We conduct the COMSOL simulation using its \textit{Semiconductor} module, where we define a radially symmetric geometry to model the top contact and its surroundings, see Extended Data Fig.~2\textbf{a}. From bottom to top, we effectively introduce the combined effect of the graphite back-gate together with the separating \textit{h}BN layer beneath the graphene by defining the bottom surface of the whole geometry as a \textit{Thin Insulator Gate}. Thus, the back-gate voltage $V_G$ can be applied to the system. To the bulk of the lowest layer, we assign a material \textit{C} with electron affinity defined as $4.6\,\mathrm{V}$ 
 in accordance with measured values of graphene~\cite{yu_tuning_2009,shi_work_2010,ziegler_variation_2011}. The
linear dispersion of graphene is introduced by defining the density of states in the
valence and conduction bands as $n=V_G\cdot 2.72\times10^{12}\  \text{m}^{-2}$.

The graphene layer is covered by two separate parts from the top. The metal contact is defined as the frustum-shaped area in the center with material type \textit{Air} and \textit{Metal Contact} interface to the top surface of the graphene layer. For the latter, we use the work function of chromium $W_\mathrm{Cr} = 4.3\,\mathrm{V}$~\cite{song_determination_2012}. The top peripheral area pertains to the \textit{h}BN layer that encapsulates the graphene
from the top. Its material is set as \textit{$SiO_2$}, whereas the work function
and the band gap are adjusted for \textit{h}BN.
\subsection*{Effective model for the emergent parabolic potential}
We model the interpolating electrostatic potential $\phi$ using a parabolic onsite function $V(r, V_b)$ on a tight-binding lattice, where the Fermi level E = 0 is defined with respect to the
primitive work function of graphene. The onsite potential is a piece-wise function
\begin{equation}\label{eq: V_alpha}
    V(r,V_b)=\begin{cases}
    V_0 & r<0\\
    V_b+\text{sign}(V_0-V_b)\alpha(r-r_0)^2 & 0\leq r<r_0\\
    V_b & r\geq r_0,
    \end{cases}
\end{equation}
where $V_0=-0.35t$ and $r_0=\sqrt{|V_b-V_0|/\alpha}$. The smoothness of $V$ is defined as $\alpha=0.000027\,t/\text{site}$ (we omit the unit of $\alpha$ in the following discussion), such that the electrostatic potential obtained by COMSOL is best fitted around the cavity region,
see Extended Fig.~2\textbf{d}. We define the hopping amplitude $t=0.46\ \mathrm{eV}$ to model a $500\ \mathrm{nm}$ system with $301$ sites. 
\subsection*{Quasi-1D model}
The tight-binding Hamiltonian of graphene system reads
\begin{equation}
\begin{split}
    H_{2D}=t\sum_{m,n}\big(&c_{A,m,n}^\dagger c_{B,m,n}+c^{\dagger}_{B,m+1,n}c_{A,m,n}\\
    &+c^{\dagger}_{B,m,n+1}c_{A,m,n}+c.c.\big),
\end{split}
\end{equation}
where $c^\dagger_{l,m,n}$ denotes an electron creation operator on the sublattice $l=A$$(B)$ at site $(m,n)$. Next, as the interface to the contact is a closed circle, we impose a cylindric geometry, i.e. the boundary in $n$-direction is closed with periodic boundary conditions such that we can reduce the 2D system to a collection of 1D channels labeled by the good quantum number $k_\perp$. The left end of the cylinder corresponds to the interface to the contact(Extended Data Fig.~3\textbf{a}). Moreover, we introduce the potential profile in Eq.~\eqref{eq: V_alpha} modeling the electrostatic potential along the $\hat{r}$ direction. The corresponding Hamiltonian is given as
\begin{equation}\label{eq:tight-binding 1D}
\begin{split}
    H_{k_\perp}&=t\sum_{m}\big(c_{A,m,k_\perp}^\dagger c_{B,m,k_\perp}\\  &+c^{\dagger}_{B,m+1,k_\perp}c_{A,m,k_\perp}+e^{i\sqrt{3}k\perp}c^{\dagger}_{B,m,k_\perp}c_{A,m,k_\perp}+c.c.\big)\\
&+\sum_{m}\sum_{l=A,B}V(m,V_b)c^{\dagger}_{l,m,k_\perp}c_{l,m,k_\perp},
\end{split}
\end{equation}
where $m=1,\dots,M$ labels the site along $r$-direction. The spectrum is obtained using exact diagonalization as a function of $k_\perp$ with $V_b=0$, see Extended Data Fig.~3\textbf{b}. We find that there are discrete spectrum flows existing next to the continuous spectrum. While the latter represents the continuous energy levels in the graphene bulk, the former indicates the existence of the discretized modes confined in the cavity.

\subsection*{Local density of states and conductance}
We use the Green's function method with the 1D system in Eq.~\eqref{eq:tight-binding 1D} to calculate the local density of states (LDOS) and the transport across the system. The local density of states at site $m$ and at the Fermi surface $E=0$ is given as~\cite{Datta_1995}
\begin{equation}\label{eq: LDOS GF}
    \text{LDOS}_{k_\perp}(m)=-\frac{1}{\pi}\mathrm{Im}[G^R_{k_\perp}(m,m;0)]\,,
\end{equation}
where $G^{R}_{k_\perp}(m,m';E)$ is the retarded Green's function between site $m$ and $m'$ at energy $E$. As the system is coupled to leads at the two ends, the retarded and advanced Green's functions are modified by the interaction between the system and leads as
\begin{equation}
\begin{split}
    G^{R}_{k_\perp}(m,m';E)&=[E-H_{k_\perp}-\Sigma^R_l-\Sigma^R_r]^{-1}_{m,m'},\\
    G^{A}_{k_\perp}(m,m';E)&=[E-H_{k_\perp}-(\Sigma^R_l)^\dagger-(\Sigma^R_r)^\dagger]^{-1}_{m,m'},
\end{split}
\end{equation}
where $\Sigma^R_{l/r}=$ is the self-energy due to the leads. We calculate the LDOS for each site $m$ as a function of $V_b$ and $k_\perp$. To illustrate the discrete energy levels in the cavity, we average over its extent$\text{DOS}_c=\sum_{m=1,\dots30}\text{LDOS}(m)$, i.e. we sum over the LDOS of the first 30 sites  ($\approx50\,\mathrm{nm}$) next to the contact. See in Extended Data Fig.~3\textbf{c}.

Next, the $k_\perp$ conductance between the two leads is $G_{k\perp}=e^2T_{k\perp}/h$, where the transmission probability $T_{k\perp}$ is obtained as
\begin{equation}\label{eq: conductance GF}
T_{k_\perp}=\mathrm{Tr}\left[\Gamma_lG^R_{k_\perp}\Gamma_rG^A_{k_\perp}\right],
\end{equation}
where $\Gamma_{l/r}=i[\Sigma^R_{l/r}-(\Sigma^R_{l/r})^\dagger]$. We numerically evaluate Eq.~\eqref{eq: conductance GF} as a function of $k_\perp$ and $V_b$, see Extended Data Fig.~3\textbf{d}.
To compare with the experiments, we introduce the carrier density $n_s$ using the relation $n_s=(V_bt)^2/(\pi v_F^2\hbar^2)$, where $v_F=10^6\,\mathrm{m/s}$ is the Fermi velocity of the graphene.

\subsection*{Analytic expression for the cavity Spectrum and tunneling strength $t_{pn}$}
We calculate the spectrum of the cavity mode using the WKB approximation. For the mode with transverse momentum $k_\perp$ in a cavity of length $l_c$, we obtain the 
the Sommerfeld quantization condition by requiring the continuity of the wave function at the boundary $r=l_c$~\cite{zalipaev_spectrum_2013} 
\begin{equation}\label{Eq: Quantization condiction}
\begin{split}
   \mathrm{Re}\bigg(2\int_{0}^{l_c}\mathrm{d}r\frac{1}{v_\parallel}\sqrt{[E-V(r,V_b)]^2-v_\perp^2|k_\perp|^2}&\bigg)\\
   =(2m-&\frac{1}{2})\pi, 
\end{split}
\end{equation}
where $m\in\mathbb{N^+}$ is an integer number characterizing the quantized energy levels. To cope with the 1D Hamiltonian in Eq.~\eqref{eq:tight-binding 1D}, we chose the Fermi velocity $v_\parallel=1$ and $v_\perp=3/2$ for the energy in the unit of $t$. To study the DOS inside the cavity, we solve $V_b$ as a function of $k_\perp$ for $E=0$ and $m\in\mathbb{N}^{+}$ using the equation above, i.e. for a mode confined in the cavity characterized by the momentum $k_\perp$, it can be found at the Fermi level $(E=0)$ when the graphene is doped to $V_b$. As such, we obtain an effective spectrum $\epsilon_c(m,k_\perp)$ for the cavity mode, where $\epsilon_c(m,k_\perp)\equiv V_b$ for given quantized number $m$ and momentum $k_\perp$.

We numerically evaluate the integral in Eq.~\eqref{Eq: Quantization condiction} and show the obtained $\epsilon_c(m,k_\perp)$ as the white dashed line in Extended Data Fig.~3\textbf{c}. The analytica results agree well with the LDOS obtained numerically.

Finally, we estimate the transmission amplitudes $t_{pn}$ using the WKB approximation \cite{shytov_klein_2008}:
\begin{equation}\label{eq:wkb t}
t_{pn}=e^{-\mathrm{Im}\int^{l_c+l_f}_{l_c}k_\parallel(r)\mathrm{d}r},
\end{equation}
where the imaginary wave-vector $k_\parallel=\sqrt{V_\alpha(r,V_b)^2-v_F^2|k_\perp|^2}/v_F$ characterizes the exponentially decaying wave function in the classically forbidden region with a length $l_f=\sqrt{(V_b+v_F|k_\perp|)/\alpha}-\sqrt{(V_b-v_F|k_\perp|)/\alpha}$. In Extended Data Fig.~3\textbf{e}, we show $t_{pn}$ obtained by numerically evaluating Eq.~\eqref{eq:wkb t} as a function of the density carrier $n_s$ and the momentum $k_\perp$.

\renewcommand{\figurename}{Extended Data Fig.}
\setcounter{figure}{0}

\section*{Extended data figures and tables}

\begin{figure*}[h]
		\centering
		\includegraphics[width=13.5cm]{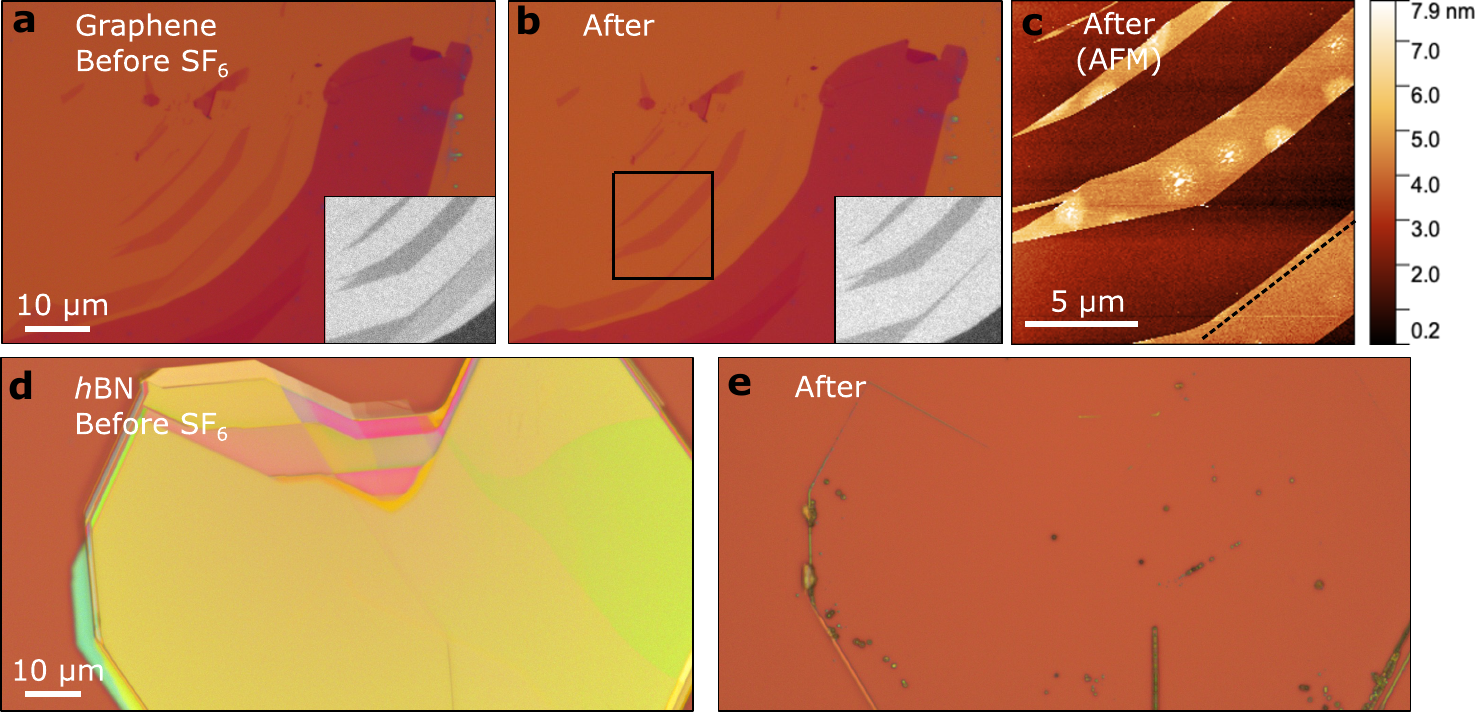}
		\caption{\textbf{- Selective etching.} (\textbf{a}-\textbf{b}) Optical images of a cluster of graphene flakes down to monolayer thickness before and after 60\,s of SF$_6$ plasma etching, 80\, mTorr, 30\,W, 80\,sccm. Insets show enhanced contrast optical images, showing monolayer graphene with reduced optical contrast after processing. \textbf{c.}  Atomic force microscope (AFM) scan of the graphene after processing, highlighting the continued presence of monolayer graphene. (\textbf{d}-\textbf{e}) Optical images of \textit{h}BN before and after processing, demonstrating the high selectivity between the graphene and \textit{h}BN.}
    \label{fig:ext1}
\end{figure*}

\begin{figure*}[!t]
		\centering
		\includegraphics[scale=1]{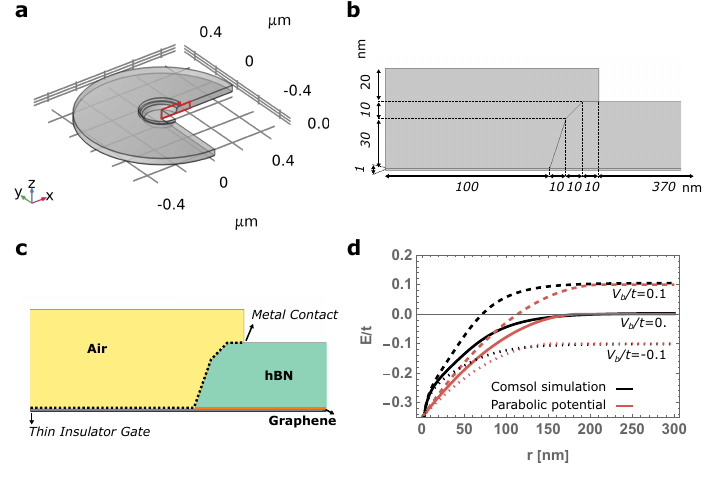}
		\caption{\textbf{- COMSOL simulation of the metal-on-graphene device.} \textbf{a.} To model the top contact and its surroundings, we define a radially-symmetric shape with total radius of $500\,\textrm{nm}$. \textbf{b.} Side view zoom-in on the red line in \textit{a}, detailing the dimensions of different parts in the contact region. \textit{c.} We use the \textit{Semiconductor} module of COMSOL to define the different material properties in the junctions and their contacts.\textit{d.} The onsite potential modeled by a parabolic function $V(r,V_b)$, compared with the simulation data from COMSOL. We chose the parameters $\alpha=0.000027$ and $V_0=-0.35t$ such that the length of the interpolation region matches with the simulation data. 
  }
		\label{fig: COMSOL setup}
\end{figure*}
\begin{figure*}[ht]
		\centering
		\includegraphics[scale=1]{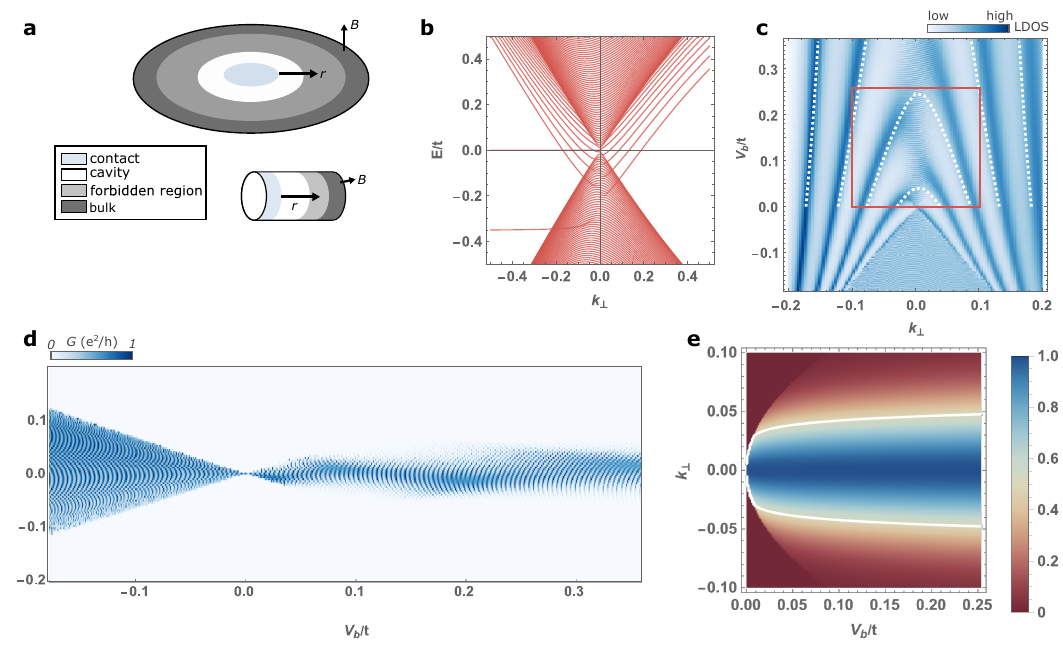}
		\caption{\textbf{a.}  The Corbino-disk shape system (upper panel) is mapped to a cylindric geometry (low panel). \textbf{b.}  The spectrum of the $1D$ system for $V_b/t=0$, obtained by solving Eq.~\eqref{eq:tight-binding 1D} in a system with $M=301$, $\alpha=0.000027$ and $V_0=-0.25t$. \textbf{c.} The DOS in the cavity as a function of the transverse momentum $k_\perp$ and $V_b$, obtained by numerically evaluating Eq.~\ref{eq: LDOS GF}. The dashed white lines indicate the dispersion of the bound state in the cavity, obtained by solving Eq.\ref{Eq: Quantization condiction}. The LDOS shown in the main text (Fig.~\ref{fig: 1D system, resistance with varying sites 2}\textbf{c}) is marked using the red solid line. \textbf{d.}  We calculate the conductance across a 1D system with the length $M=801$ as a function of $k_\perp$ and $V_b$ by evaluating Eq.~\eqref{eq: conductance GF} numerically. The parameters of $V(r,V_b)$ are the same as in \textbf{b}.\textbf{e.}  The tunneling strength $t_{pn}$ through the classical forbidden region as a function of $V_b$ and $k_\perp$, obtained by evaluating Eq.~\eqref{eq:wkb t}. The thick white line indicates $t_{pn}=0.5$.}
		\label{fig: parabolic tpn}
\end{figure*}
\begin{figure*}[ht]
		\centering
		\includegraphics[scale=1]{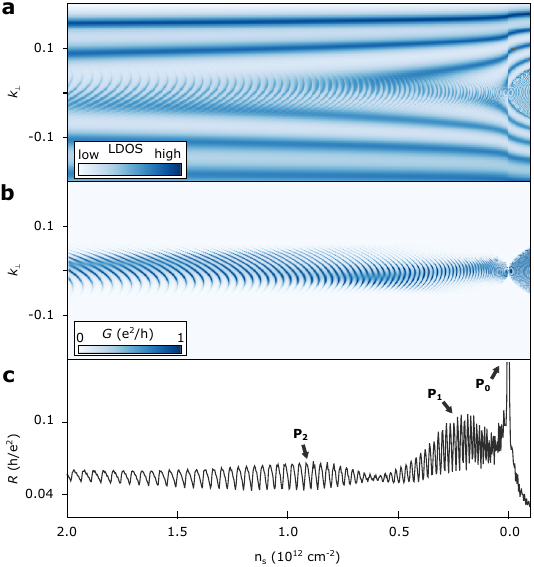}
		\caption{\textbf{- Numeric simulation for large parameter range.} \textbf{a.} The DOS in the cavity as a function of the transverse momentum $\mathbf{k}_\perp$ and carrier density $n_s$, obtained by evaluating Eq.~\eqref{eq: LDOS GF} using the same parameter as in Extended Data Fig.~3\textbf{d}. \textbf{b.} The conductance across the system, obtained by solving Eq.~\eqref{eq: conductance GF}. \textbf{c}. The corresponding resistance as a function of $n_s$. A sharp peak $\mathbf{P}_0$ and two smooth peaks $\mathbf{P}_1$, $\mathbf{P}_2$ are observed. The small oscillations appear due to the finite-size effect in the numeric simulation.}
		\label{fig: numeric large range}
\end{figure*}
\begin{figure*}[ht]
		\centering
		\includegraphics[scale=1]{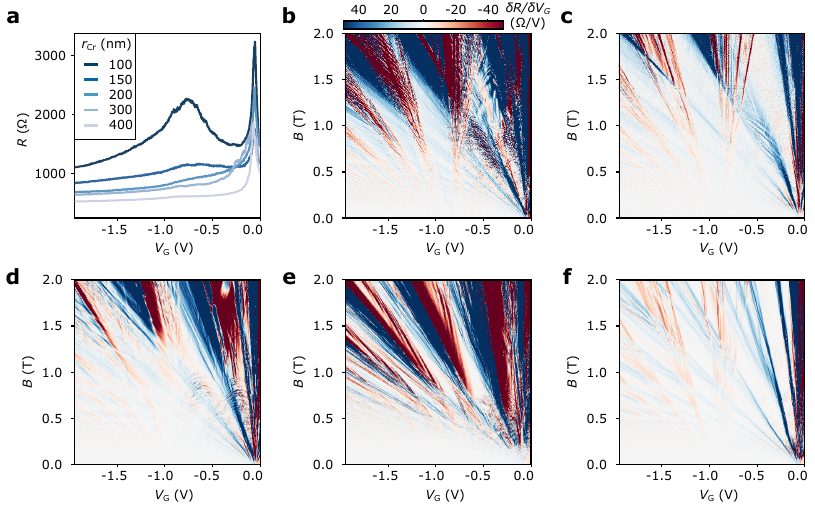}
		\caption{\textbf{- Low field measurement for different contacts.} \textbf{a.} Experimentally measured resistance at zero field as a function of the back gate voltage $V_G$. \textbf{b.}-\textbf{e.} Differential resistance at low-field and small $V_G$, measured with the contact's size $r_\mathrm{Cr}= 100, 150, 200, 300$ and $400\,\mathrm{nm}$, respectively.}
		\label{fig: LFs}
\end{figure*}
\begin{figure*}[ht]
		\centering
		\includegraphics[scale=1]{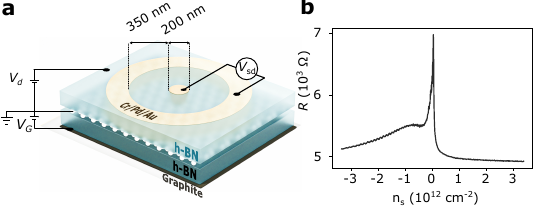}
		\caption{\textbf{- Low field measurement for Corbino disk setup.} \textbf{a}. The device is defined in the Corbino disk geometry using the same fabrication method as in Fig.~\ref{fig: setup and zero-resistance}\textbf{a}. A dot-shaped top contact with a radius $100\,\mathrm{nm}$ is placed in the middle. The edge-to-edge distance between the middle contact and the outer annular-shaped contact is $350\,\mathrm{nm}$. \textbf{b}. Two-terminal resistance between the outer contact and the small dot contact as a function of the carrier density $n_s$.}
		\label{fig: Corbino disk measurement}
\end{figure*}

\end{document}